# Product Line Metrics for Legacy Software in Practice

Christian Berger, Holger Rendel, Bernhard Rumpe
Software Engineering
RWTH Aachen University
Aachen, Germany
http://www.se-rwth.de/

Carsten Busse, Thorsten Jablonski, Fabian Wolf
Electronics Development
Volkswagen AG Business Unit Braunschweig
Braunschweig, Germany
http://www.volkswagen.de/

***Abstract*—Nowadays, customer products like vehicles do not only contain mechanical parts but also a highly complex software and their manufacturers have to offer many variants of technically very similar systems with sometimes only small differences in their behavior. The proper reuse of software artifacts which realize this behavior using a software product line is discussed in recent literature and appropriate methods and techniques for their management are proposed. However, establishing a software product line for integrating already existing legacy software to reuse valuable resources for future similar products is very company-specific. In this paper, a method is outlined for evaluating objectively a legacy software's potential to create a software product line. This method is applied to several development projects at Volkswagen AG Business Unit Braunschweig to evaluate the software product line potential for steering systems.**

## I. Introduction and Motivation

In recent literature [1], concepts for creating a software product line are outlined which discuss either top-down methodologies in developing new products according to a product line which is created from scratch, or bottom-up methodologies to refactor an existing software architecture towards a product line to serve similar products. However, guidelines are missing on how to evaluate and select suitable parts of *existing legacy* software in a practical and economic way to create a software product line.

Therefore, in [2] several metrics are elaborated to evaluate objectively a set of existing software products to gather information about their existing potential for creating a software product line. In this paper, these metrics are applied at Volkswagen AG Business Unit Braunschweig for evaluating existing steering systems which are implemented in the Volkswagen Tiguan and Passat to estimate the benefit of creating a software product line. Because these steering systems are safety-critical, a proper reuse of existing artifacts is highly desirable whose elements have to be selected by relying on an objective evaluation which is outlined in the following.

## II. Related Work

Metrics to evaluate an architecture's capabilities to form a product line are outlined by [3] which base on interface specifications and which are limited to object-oriented systems only. A similar approach about measuring object-oriented systems is presented in [4]. In that contribution the metrics base on a concept called *service utilization* which includes public method signatures and directly accessible data structures.

[5] outlines metrics for a core asset which is divided into a product line architecture, a component model, and a decision model which has a wider scope and does not consider components in detail.

Another approach using metrics is provided by [6] which can be used to evaluate the quality of a product line by using a given variability model which does not exist in our evaluation. [7] use function nets combined with feature models to provide metrics for determining the necessary effort to integrate further features and functionality into an existing product line and, thus, considerations about existing similar components are not necessary.

Modal transition systems (MTS) [8] have a similar mathematical model compared to the model which is used here. However, MTS model the behavior of a system rather than the system's architecture and thus, evaluating compositional aspects for existing legacy software artifacts to form a software product line are not possible.

## III. Relationship-based Metrics Model

To carry out an objective evaluation of already existing legacy software, a formal representation of its internal structure is required. This representation is necessary to have both a programming language-independent and a common representation to apply metrics. Thus, these metrics can be applied to various existing legacy software even in heterogeneous project contexts.

In the following, a formal relationship model is defined which is used to discover similarities in already existing legacy software artifacts with respect to a predefined *similarity threshold*. At last, this relationship model is used by different metrics to calculate various *intersection sets* for the considered set of products.

### A. Formal Representation of Internal Structure

To identify the relationship between $n$ considered software products each product $p_i$ with $1 \leq i \leq n, n \geq 2$ is decomposed into a set $\mathbb{C}_{p_i}$ of $m$ so-called *atomic* pieces $c_j$ with $1 \leq j \leq m, m \geq 1$ which are called *components* according to [9]. These components gather functionality of user visible features. Two components $c_j$ and $c_k$ are called *extrinsically equal* nonetheless to which product $p_i$ they belong to as denoted by $c_j = c_k$ iff they have the same name.

Next, the syntactical interface set $\mathbb{S}_{c_j}$ is defined for the component $c_j$. This set consists of $r$ concrete signatures $s_k = id \times \mathcal{P}(\{\mathbb{N}, \mathbb{R}, [\![\text{TYPE}]\!], \dots\})$ with $1 \leq k \leq r, r \geq 1$ where *id* denotes a unique identifier for a concrete signature. Two signatures $s_p$ and $s_q$ are called *equal* as denoted by $s_p = s_q$ iff they exhibit the same signature.

With each signature $s_k$, the behavior $b_{s_k} : s_k \rightarrow Out_k$ is associated with $Out_k = \mathcal{P}(\{\mathbb{N}, \mathbb{R}, [\![\text{TYPE}]\!], \dots\})$. The behavior $b_{s_k}$ defines for a given input vector *in* from the input set $In = \mathcal{P}(\{\mathbb{N}, \mathbb{R}, [\![\text{TYPE}]\!], \dots\})$ passed through the signature *id* the output vector *out* from the output set $Out_k$. Two behaviors $b_{s_k}$ and $b_{s_o}$ are called *equal* as denoted by $b_{s_k} = b_{s_o}$ iff $\forall in \in In : b_{s_k}(in) = b_{s_o}(in)$.

### B. Relationship Model to Discover Similarity

A *Relationship Model* $\Delta_{\mathcal{R}}$ is required to evaluate the relationship between two considered components $c_i$ and $c_j$ from two products $p_a$ and $p_b$. This model defines the operator $\Delta : \mathbb{C} \times \mathbb{C} \rightarrow [0 \dots 1]$ to describe the relationship. Using the specification for the relationship model two variants are defined: $\Delta_{\equiv}$ and $\Delta_{\approx}$. The former is called *identity* and is defined by

$$\Delta_{\equiv}(c_x, c_y) = \begin{cases} 1 \;\Leftrightarrow & c_x = c_y \;\wedge \\ & \forall s \in \mathbb{S}_{c_x} \exists t \in \mathbb{S}_{c_y} : \\ & \quad s = t \;\wedge b_s = b_t \;\wedge \\ & \forall w \in \mathbb{S}_{c_y} \exists v \in \mathbb{S}_{c_x} : \\ & \quad w = v \;\wedge b_w = b_v \\ 0 & \textit{else}. \end{cases} \tag{1}$$

$\Delta_{\equiv}$ can be used to identify components which are *syntactically* and *semantically* identical. Therefore, two considered components must be at least *extrinsically equal* and exhibit the same signature set to be evaluated.

The second *Relationship Model* is called *gradual similarity* and is defined by

$$\begin{aligned} \Delta_{\approx}(c_x, c_y) &= \frac{|W|}{\left|\mathbb{S}_{c_x} \cup \mathbb{S}_{c_y}\right|} \\ \textit{with } W &= \{s \in \mathbb{S}_{c_x} | \exists t \in \mathbb{S}_{c_y} : out_s = out_t\}. \end{aligned} \tag{2}$$

Contrary to the aforementioned relationship model, $\Delta_{\approx}$ does not require that two considered components $x$ and $y$ must exhibit the same name. Instead, the main idea behind this model is to analyze the components' behavior by trying to identify the same behavior from component $x$ in the other component $y$ which might be exported by a slightly different signature. For example, imagine two signatures which provide the behavior of a sorting algorithm for an array of numbers. Component $x$ implements a bubble sort algorithm exporting the signature `sort(bool ascending, List<int> list);` and component $y$ implements a quicksort algorithm exporting the signature `sortList(List<int> list, bool ascending);`. Both methods exhibit the same behavior but they are called in a different way.

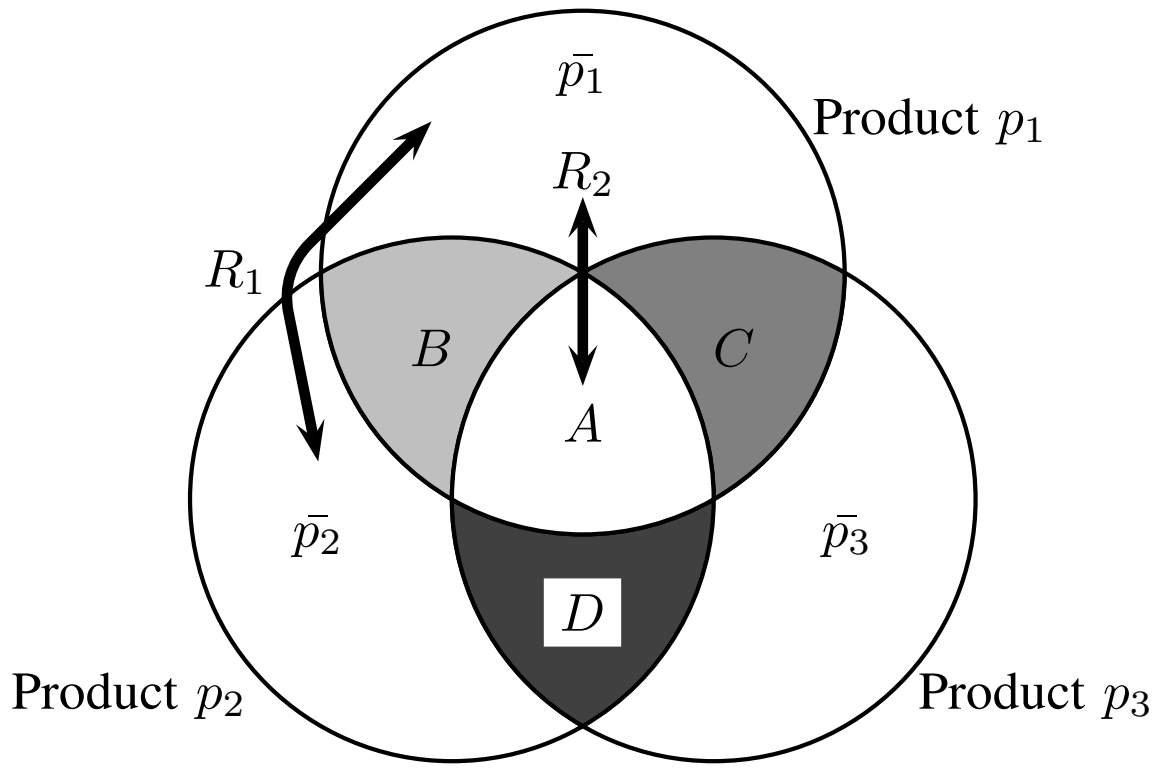

Fig. 1. Evaluation of three products $p_1$, $p_2$, and $p_3$: The circles denote the set of components for each product; $\bar{p_1}$ denotes the complementary set of components for product $p_1$ without the sets $B$, $A$, and $C$. $A$ indicates the set of components which are shared among all three products with respect to the selected relationship model $\Delta_{\mathcal{R}}$. $B$ contains all components which are shared only by $p_1$ and $p_2$; $C$ and $D$ are calculated in an analog manner. $R_1$ and $R_2$ denote different relationships.

### C. Metrics for Evaluating Legacy Software Artifacts

By defining a relationship model $\Delta_{\mathcal{R}}$ which should be immutable during the analysis of a given set of products to get comparable results, legacy software artifacts can be analyzed. Therefore, metrics as outlined in [2] are required to evaluate their potential to create a product line.

The main idea behind these metrics is to correlate various intersection sets as shown in Figure 1. These intersection sets are determined by the aforementioned relationship model. Obviously, the size of each intersection set depends directly on the selected relationship model.

In the following, the metrics which are required in Section IV are described briefly; for a comprehensive elaboration please refer to [2].

**Size of Commonality.**

$$\begin{aligned} SoC_{\Delta_{\mathcal{R}},N} &= \left|\bigcap\nolimits_{i,j=1\dots n, i<j} P_{i,j}(N)\right| \\ \textit{with } P_{i,j}(N) &= \{c \in \mathbb{C}_{p_i} | \exists c' \in \mathbb{C}_{p_j} : \Delta_{\mathcal{R}}(c, c') \geq N\}. \end{aligned} \tag{3}$$

In Equation (3), the *Size of Commonality* is defined which is calculated from set $A$ in Figure 1 containing the number of components which are shared among all considered products. Therefore, the calculated relationship model $\Delta_{\mathcal{R}}$ must be greater than a predefined similarity threshold $N$. The components from this set are used in each product and, thus, have the greatest reusability.

**Product-related Reusability.**

$$\begin{aligned} PrR_{\Delta_{\mathcal{R}},N,i} &= \frac{SoC_{\Delta_{\mathcal{R}},N}}{\left|\mathbb{C}_{p_i}\right|} \\ \textit{with } \left|\mathbb{C}_{p_i}\right| &> 0. \end{aligned} \tag{4}$$

The ratio in Equation (4) calculates the reusability of all commonly available components related to a specific product $p_i$: The greater the ratio the better is the reusability of all commonly shareable components. This ratio is described by class $R_2$ in Figure 1.

**Individualization Ratio.**

$$\begin{aligned} IR_{\Delta_{\mathcal{R},N},i} &= \frac{|W_{\Delta_{\mathcal{R},N},i}|}{|\mathbb{C}_{p_i}|} \\ with\ W_{\Delta_{\mathcal{R},N},i} &= \{c \in \mathbb{C}_{p_i} | \forall c' \in \mathbb{C}_{p_k} : \Delta_{\mathcal{R}}(c,c') < N \\ & \quad for\ k = 1 \ldots n \wedge k \neq i\}, \\ & \quad |\mathbb{C}_{p_i}| > 0. \end{aligned} \tag{5}$$

Equation (5) calculates the product-related *Individualization Ratio* to describe the product's specific individualization related to the amount of components which are shared with other products. Obviously, the smaller this ratio the greater is the product's benefit to reuse other components.

The aforementioned metrics are only a brief selection from [2] which are mainly used to evaluate the considered product set at Volkswagen AG Business Unit Braunschweig.

## IV. Case Study at Volkswagen AG Business Unit Braunschweig

We applied the aforementioned relationship models alongside the metrics outlined above to three steering systems which are realized with MATLAB/Simulink to provide a decision support for the software architects to estimate the effort which is necessary to establish a software product line from the existing products. Therefore, we first evaluated the already existing potential of the considered set of products to create a product line for the assumption of avoiding any modifications to the existing components i.e. to create a software product line without changing any component and to reuse only syntactically and semantically identical components. This evaluation was carried out using the $\Delta_{\equiv}$ relationship model.

Then, we changed the relationship model to allow gradual comparisons with various similarity thresholds using $\Delta_{\approx}$. We calculated the metrics which are outlined above for these similarities to evaluate the different potential to create a software product line.

The results are the basis for a benefit-cost analysis to estimate the required effort for establishing a software product line with a desired potential.

### A. Evaluating the Current Potential of the Existing Products

Starting with the analysis of the current potential of the existing set of products the relationship model $\Delta_{\equiv}$ as defined in Equation (1) was applied to evaluate the products' components. Thus, only components which are syntactically and semantically identical are considered to be shared among all considered products.

TABLE I
Evaluating the currently existing set of products using $\Delta_{\equiv}$.

| $\|\bar{p_1}\|$ | $\|\bar{p_2}\|$ | $\|\bar{p_3}\|$ | $\|A\|$ | $\|B\|$ | $\|C\|$ | $\|D\|$ |
|---|---|---|---|---|---|---|
| 21% | 25% | 23% | 13% | 4% | 9% | 4% |

In Table I the results for the three steering systems based on Figure 1 are shown; for preserving the intellectual property only percentage quotations are provided. The steering system $p_1$ consists of 47% of all components, steering system $p_2$ consists of 49% of all components, and the last steering system $p_3$ contains 51% of all components. In the table the residual sets for each product's component which are only product-specific are depicted in the first three columns; for example, the residual set for $p_1$'s components is denoted by $\bar{p_1}$. Next, the size of $A$ which represents the size of commonality as defined by Equation (3) is shown. The last three columns represent the pairwisely calculated intersection sets as shown in Figure 1.

TABLE II
Applying the metrics for $\Delta_{\equiv}$.

| $PrR_{\Delta_{\equiv},1}$ | $PrR_{\Delta_{\equiv},2}$ | $PrR_{\Delta_{\equiv},3}$ |
|---|---|---|
| 0.27 | 0.26 | 0.25 |

| $IR_{\Delta_{\equiv},1}$ | $IR_{\Delta_{\equiv},2}$ | $IR_{\Delta_{\equiv},3}$ |
|---|---|---|
| 0.45 | 0.52 | 0.46 |

As shown in Table II nearly half of the three products' components are product-specific as shown by the ratio *IR*. Moreover, the product-related reusability of components which are shared among all considered products is nearly 25% on average. These results reflect the potential of a software product line when all considered products' components are participating in an unmodified manner. Thus, a software product line which consists of syntactically and semantically components would save resources at least for maintenance reasons of these commonly used components.

Next, an evaluation is of substantial interest where components of the considered set of products can be modified to increase their reusability of existing shared components to optimize the benefit of creating a software product line.

### B. Changing the Relationship Model to Increase Similarity

Another relationship model must be chosen to evaluate the potential of the considered set of products' components to be part of a newly created software product line when these components are allowed to be modified slightly to increase their reusability. Therefore, the relationship model $\Delta_{\approx}$ as defined in Equation (2) is used for different similarity thresholds.

TABLE III
Evaluating the set of products using $\Delta_{\approx}$ for different thresholds.

| N | $\|\bar{p_1}\|$ | $\|\bar{p_2}\|$ | $\|\bar{p_3}\|$ | $\|A\|$ | $\|B\|$ | $\|C\|$ | $\|D\|$ |
|---|---|---|---|---|---|---|---|
| 0.99 | 50% | 74% | 50% | 10% | 0% | 12% | 2% |
| 0.95 | 50% | 65% | 50% | 14% | 0% | 8% | 2% |
| 0.90 | 45% | 61% | 46% | 17% | 0% | 8% | 2% |
| 0.85 | 41% | 57% | 42% | 20% | 0% | 9% | 2% |
| 0.80 | 36% | 52% | 33% | 20% | 0% | 11% | 5% |
| 0.75 | 32% | 48% | 29% | 24% | 0% | 12% | 5% |
| 0.70 | 32% | 43% | 29% | 27% | 0% | 10% | 5% |
| 0.65 | 27% | 43% | 29% | 30% | 0% | 10% | 3% |
| 0.60 | 23% | 39% | 25% | 34% | 0% | 11% | 3% |
| 0.50 | 18% | 39% | 21% | 35% | 0% | 14% | 3% |
| 0.40 | 18% | 30% | 17% | 40% | 0% | 11% | 6% |
| 0.30 | 14% | 22% | 13% | 42% | 3% | 12% | 9% |
| 0.15 | 9% | 13% | 8% | 53% | 3% | 10% | 1% |
| 0 | 5% | 4% | 4% | 67% | 4% | 7% | 1% |

In Table III an analogous representation as already shown in Table I is depicted. Here, the first column represents the gradual relationship between 0%–99%; the following columns describe the amount of components like the aforementioned table. Contrary to the relationship model $\Delta_{\equiv}$, $\Delta_{\approx}$ also allows to compare signatures which does not need to be equal. Thus, these ratios are different from Table II.

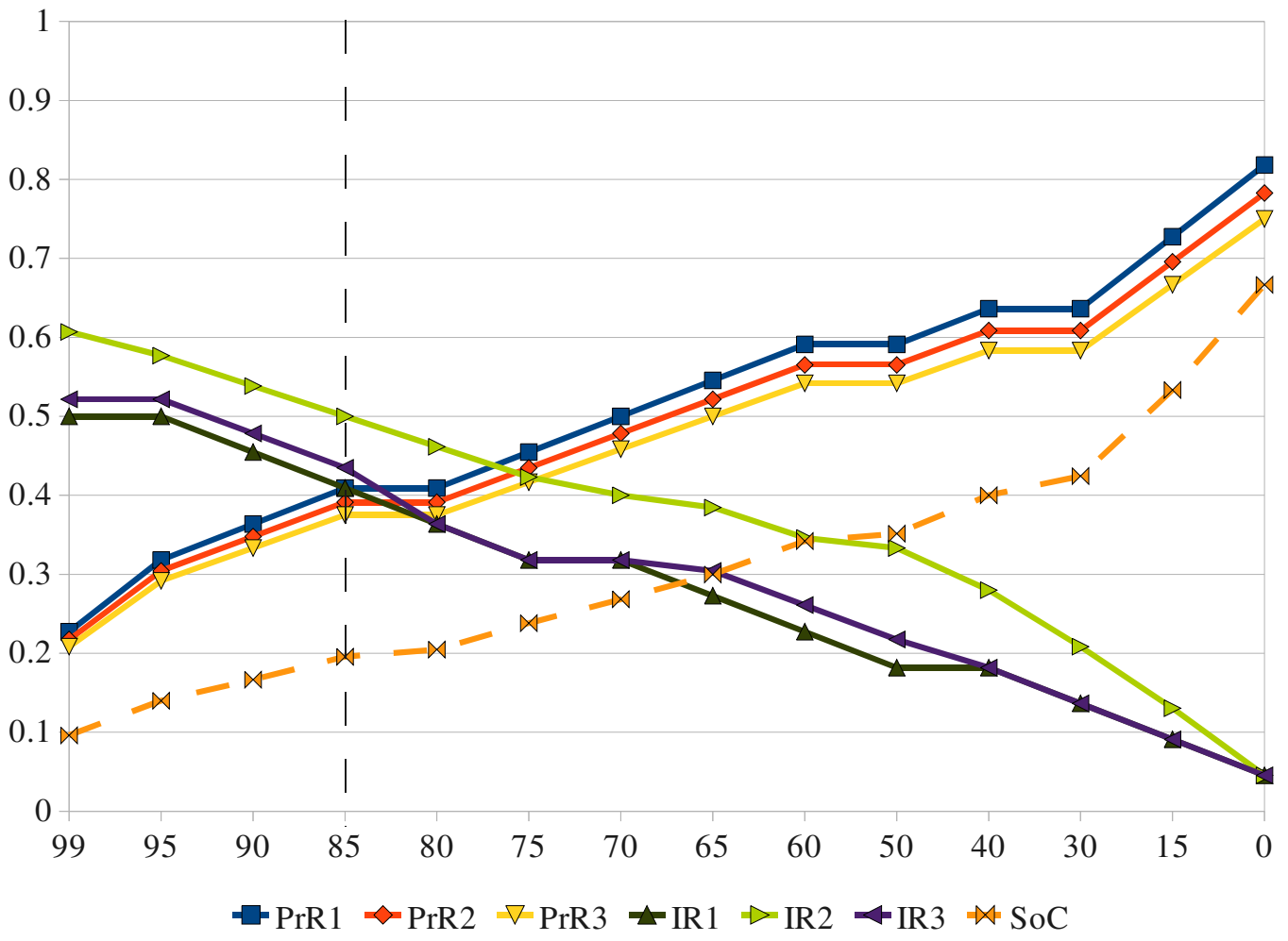

Fig. 2. Applying the metrics to the various relationship models: On the X-axis the gradual relationship 99%–0% is shown; on the Y-axis the ratios *PrR*, *IR*, and *SoC* for all three considered products are drawn.

In Figure 2 the aforementioned metrics are applied to the products' components which are compared using the gradual relationship model $\Delta_{\approx}$. Obviously on one hand, all ratios for *IR* are decreasing because the relationship model assumes to refactor the existing components to increase their benefit of reusing shared components; on the other hand, the corresponding *PrR*s are increasing analogously.

As already stated in Section IV-A, a software product line for the existing set of products is valuable. Using Figure 2, further effort in modifying the existing components is desirable to increase the benefit for the current products' components. For example, if only 15% ($N = 85\%$) of all components' behavior is modified the reusability of these components can be increased and the *SoC* can be enlarged by more than 50% related to the relationship model $\Delta_{\equiv}$. This is also the ratio where a product's reusability benefit is greater than its individualization ratio; this point can be achieved by adjusting only 15% of a component's behavior. Thus, maintenance tasks as well as innovations for commonly shared components can be more easily carried out for the existing set of products.

Contrary to the creation of a software product line for unmodified components, resources are required to adjust these existing components. However, these required resources depend from a concrete project's context. Thus, the metrics and relationship model which are outlined in this paper provide an objective decision support for the software architects and project managers to estimate the necessary effort to create of software product line with a certain benefit as well as to realize a business-specific reusability strategy.

## V. Conclusion

This contribution reports on an examination of three rather individually developed product variants consisting of safety-critical software components from a set of steering systems at Volkswagen AG Business Unit Braunschweig and evaluates their ability to form a product line using a relationship-based metrics model. The results show that potential for a software product line exists from an objective point of view for different ways of creating the software product line. These results enable the software architects and project managers to estimate the required effort to the setup a software product line to achieve a desired reusability effect.

Further elaboration will be carried out to extend the gathered results from the relationship models. One aspect which is currently of substantial interest is to combine the gradual similarity model with methods to estimate more objectively the required effort to modify the considered set of products accordingly to achieve a desired reusability benefit. Therefore, methods like the function point analysis or the constructive cost model (COCOMO) will be analyzed to combine them with the results from the relationship models for the applied metrics.